\documentclass[english,a4paper]{article}
\usepackage{dcolumn}
\usepackage{bm}
\usepackage{graphicx}
\usepackage{amsmath}
\usepackage{amsfonts}
\usepackage{amssymb}
\usepackage{amsthm}
\usepackage{amstext}
\usepackage{amsbsy}
\usepackage{amsopn}
\usepackage{amscd}
\usepackage{amsxtra}

\def\phi{\varphi}
\def\epsilon{\varepsilon}

\everymath{\displaystyle}

\begin{document}

\title{Reply to the comment on the letter ``Geometric Origin of the Tennis Racket Effect"}
\author{P. Marde\v si\'c\footnote{Institut de Math\'ematiques de Bourgogne - UMR 5584 CNRS, Universit\'e de Bourgogne,
	9 avenue Alain Savary, BP 47870, 21078 DIJON, FRANCE, pavao.mardesic@u-bourgogne.fr}, G. J. Gutierrez Guillen\footnote{Institut de Math\'ematiques de Bourgogne - UMR 5584 CNRS, Universit\'e de Bourgogne,
	9 avenue Alain Savary, BP 47870, 21078 DIJON, FRANCE}, D. Sugny\footnote{Laboratoire Interdisciplinaire Carnot de
Bourgogne (ICB), UMR 6303 CNRS-Universit\'e Bourgogne, 9 Av. A.
Savary, BP 47 870, F-21078 Dijon Cedex, France, dominique.sugny@u-bourgogne.fr}}

\maketitle

\begin{abstract}
The author of the comment~[arXiv:2302.04190] criticizes our  published results in Phys. Rev. Lett. \textbf{125}, 064301 (2020) about the Tennis Racket Effect (TRE). The TRE is a geometric effect which occurs in the free rotation of any asymmetric rigid body. We explain why the criticism of this comment is not valid.
\end{abstract}


The author of the comment~\cite{comment} and of the work~\cite{work} criticizes our  published results in \cite{mardesic:2020} and \cite{vandamme:2017}
about the Tennis Racket Effect (TRE). Our studies were based on a preliminary analysis by Cushman and co-workers~\cite{MSA91,cushman}. Even if it is not described in standard textbooks of classical mechanics~\cite{Goldstein50,arnold,Landau}, TRE is a very well-known phenomenon with  many videos illustrating this effect in different rigid bodies. The TRE is a geometric effect which occurs in the free rotation of any asymmetric rigid body. This effect is manifested in a tennis racket by an almost $\pi$-twist of the head of the racket when the handle performs a full rotation. We show in~\cite{mardesic:2020,vandamme:2017} that a perfect TRE can be observed in the limit of an ideal asymmetric rigid body when the moments of inertia are far from each other. We recently investigate the dynamical role of physical constraints on the moments of inertia and we find in which conditions a perfect TRE can be performed within such constraints~\cite{gutierrez:2023}.

In the comment~\cite{comment}, the author mentions that (verbatim only adapting the numbering of the references): ``In the recent work~\cite{mardesic:2020}, authors introduced and discussed a first-order differential equation that relates two of the
Euler angles obeying the Euler-Poisson equations under very special initial conditions, see Eq.~(1) in~\cite{mardesic:2020}. Since it
was obtained from the equations, that can be used to study the dynamics of a free asymmetric rigid body, authors
assumed that their equation is suitable for describing some effects in the theory of a rigid body, including the Tennis
Racket, Dzhanibekov and Monster Flip effects. However, this assumption is not justified, and the relationship of their
solution $\psi(\phi)$ to Eq.~(1) in~\cite{mardesic:2020} with the motions of a rigid body is not clear."

If we understand properly the comment~\cite{comment}, the core of the criticism is based on the choice of the moving frame, where the \emph{preserved} angular momentum $\mathbf{m}$ is taken in the direction of the $Z$-axis of the space-fixed frame (laboratory frame). The free rotation of the rigid body is described by the \emph{relative} position of two frames: the fixed laboratory frame and the moving frame attached to the rigid body. This position is described by three Euler angles whose dynamics can be derived from Euler equations. The use of such equations in studying the rotation of the rigid body is by no means our contribution to the subject, but goes back to Euler and appears in all standard textbooks of classical mechanics~\cite{Goldstein50,Landau}. In our analysis, we consider the convention where the fixed frame is the one attached to the body and the moving frame is the laboratory one. The TRE is then described in our papers by considering specific trajectories of the angular momentum close to the separatrix between the rotating and the oscillating trajectories. Our main contribution was to express the TRE in terms of elliptic integrals of the first and third kinds~\cite{vandamme:2017}. A geometric analysis is proposed in~\cite{mardesic:2020} to show that the geometric origin of TRE is a pole of a Riemann surface defined in a complex phase space. We also confirm the TRE from a numerical integration of the differential system.

The author of~\cite{comment} further claims that solutions to the Euler equations with the convention of the angular momentum $m$ in the direction of the $Z$-axis do not present the TRE. However, this statement is not true. The proof is based on the wrong assumption that the initial condition for the angular momentum is of the form $(0,0,m_3)$ in the body-fixed frame, whereas this condition is only valid in the laboratory frame. This leads to an erroneous counter-example in~\cite{comment} which in no way proves the absence of TRE.

\end{document}